\documentclass[prl,aps,showpacs,twocolumn,preprintnumbers,
amsmath,amssymb,superscriptaddress,longbibliography]{revtex4-1}
\usepackage[english]{babel}
\usepackage{amsmath,amssymb,amsfonts}
\usepackage{mathrsfs}
\usepackage{graphicx}
\usepackage[colorlinks=True,linkcolor=red,citecolor=blue,urlcolor=blue]{hyperref}
\usepackage{bookmark}
\usepackage{xcolor}
\usepackage{chngcntr}
\usepackage{braket}
\usepackage{nicefrac}
\usepackage{bm}
\usepackage{bbm}
\usepackage{braket}
\usepackage{slashed}
\usepackage[normalem]{ulem}
\usepackage{array}
\newcolumntype{C}{>{$}c<{$}}

\def\id{\mathbbm{1}}

\renewcommand{\Re}{\mathop{\mathrm{Re}}}

\renewcommand{\i}{\mathop{\mathrm{i}}}

\newcommand{\bolds}[1]{\boldsymbol #1}
\newcommand{\scr}[1]{\mathscr #1}
\allowdisplaybreaks
\DeclareMathAlphabet{\zc}{OT1}{pzc}{m}{it}

\begin{document}

\title{Non-Hermitian chiral anomalies in interacting systems}
\author{Sharareh Sayyad}
\email{sharareh.sayyad@mpl.mpg.de}
\affiliation{Max Planck Institute for the Science of Light, Staudtstra\ss e 2, 91058 Erlangen, Germany}
%

\begin{abstract}
The emergence of chiral anomaly entails various fascinating phenomena such as anomalous quantum Hall effect and chiral magnetic effect in different branches of (non-)Hermitian physics. While in the single-particle picture, anomalous currents merely appear due to the coupling of massless particles with background fields, many-body interactions can also be responsible for anomalous transport in interacting systems. In this Letter, we study anomalous chiral currents in systems where interacting massless fermions with complex Fermi velocities are coupled to complex gauge fields. Our results reveal that incorporating non-Hermiticity and many-body interactions gives rise to additional terms in anomalous relations beyond their Hermitian counterparts. We further present that many-body corrections in the subsequent non-Hermitian chiral magnetic field or anomalous Hall effect are nonvanishing in nonequilibrium or inhomogeneous systems.  
Our results advance efforts in understanding the anomalous transport in interacting non-Hermitian systems.
\end{abstract}
\date{\today}
\maketitle

\paragraph*{ Introduction.}

The chiral anomaly emerges due to the violation of classical chiral symmetry by quantum fluctuations in odd spatial dimensions. This quantum anomaly has given rise to a plethora of exotic phenomena, including anomalous decay of neutral pion in high-energy physics~\cite{Bell1969, Adler1969, Adler1969b, Bilal2008, Zinn2019}, anomaly-induced charges in baryons in quantum chromodynamics~\cite{Eto2012}, anomalous transport in condensed matter physics~\cite{Zyuzin2012, Hosur2013, Landsteiner2016, Frohlich2018, Araki2019, Arouca2022, Frohlich2023}, and the magnetic helicity transfer in the early universe due to chiral asymmetry in cosmology~\cite{Boyarsky2012, Brandenburg2017}. Aside from deepening our understanding and their experimental realizations~\cite{Huang2015, Hirschberger2016, Zhang2016, Li2016, Ong2021}, chiral anomalies are proposed to be used in advancing quantum computing, e.g., as chiral qubits~\cite{Kharzeev2019, Babaev2023}.

While chiral anomaly in condensed matter physics usually describes noninteracting massless fermions under electromagnetic fields, generalizations of this theory allow incorporating Weyl node-mixing terms~\cite{Raines2017} and treating short-range interactions between fermions~\cite{Kikuchi1992, Rylands2021, Parhizkar2023}. The latter unveils novel contributions to the anomaly equations through its nonperturbative formulation. This differs from the traditional convention of perturbative treatment of interactions in high-energy physics to explore the chiral anomaly~\cite{Adler1969b}. These perturbative studies reported the cancellation of higher-order corrections upon respecting Lorentz and chiral symmetries~\cite{Giuliani2021} and concluded the universality of chiral anomaly. As Lorentz and chiral symmetries are usually broken in interacting condensed matter systems~\cite{Raines2017, Giuliani2021, Rylands2021}, violating the chiral symmetry by many-body interactions in the absence of background fields opens new directions to explore chiral anomaly in interacting lattice models~\footnote{We note that all reported additional terms in Ref.~\cite{Kikuchi1992, Rylands2021, Parhizkar2023} originated from many-body corrections in the anomaly equation can be written as total derivatives of physical quantities.}.

Condensed matter systems with different constituent particles, e.g., electrons and phonons, can be studied as closed or open quantum systems. While in the framework of closed systems, all degrees of freedom are treated self-consistently, the formulation of open quantum systems takes advantage of tracing out some degrees of freedom with the expense of losing unitarity. It has been shown that the path-integral formulations for open or closed systems can provide an effective non-Hermitian description for these systems~\cite{Kamenev2005, Vanleeuwen2005, Sieberer2016, Michishita2020, Zirnstein2021, GomezLeon2022, Gal2023}. Here non-Hermiticity originates from the dissipative nature of open quantum systems or the imaginary parts of self-energies accounting for interactions between different subsystems in closed quantum systems.

Upon constructing non-Hermitian models, the underlying physics of open/closed systems can be unraveled using methods in non-Hermitian physics and their unique properties with no counterparts in Hermitian physics~\cite{Zyuzin2019, Ashida2020,  Wang2021, Bergholtz2021, Okuma2023}.
The emergence of defective~\cite{Sayyad2021, Ding2022, Sayyad2022ep} and non-defective~\cite{Zhang2021b, Sayyad2022pro, Sayyad2022b} degeneracies and the occurrence of exotic boundaries modes~\cite{Zhang2022} exemplify the fascinating features of non-Hermitian models. While most studies focus on exploring noninteracting systems, investigating non-Hermitian many-body physics and their dynamics have gained momentum in recent years~\cite{Fukui1998, Buca2020, Zhang2021, Nakagawa2021, Hyart2022, Yoshida2023, Sayyad2023, Mak2023}. Despite these studies, the rich transport properties of interacting non-Hermitian models are mainly unexplored. Addressing the anomalous chiral response of interacting Dirac fermions in non-Hermitian models is the primary goal of this work.

In this letter, we explore the chiral anomaly in $(1+1)$ and $(3+1)$ dimensions for non-Hermitian Dirac fermions with complex Fermi velocities coupled to complex background gauge fields. By introducing a unified notation, we bring the non-Hermitian model and its symmetrized version under the same umbrella, enabling us to identify purely non-Hermitian contributions in anomalous currents. We further present the physical consequences of the non-Hermitian chiral anomaly in non-Hermitian systems and discuss plausible platforms to realize them.

\paragraph*{ Non-Hermitian chiral anomaly in many-body systems.}
We consider a non-Hermitian model of interacting massless fermions $\Psi$, with complex Fermi velocities, in the presence of non-Hermitian gauge fields~$(V, W)$ in even $d$ dimensions. To facilitate later comparison of our non-Hermitian results with previous Hermitian calculations, we employ a notation, shown by a tilde, which unifies non-Hermitian~($\rm nh$) and its symmetrized form, a.k.a., Hermitionized~($\rm h$), models~\cite{Sayyad2022ca}.

 In this notation, the partition function~($\cal Z$), the Euclidean-space action~($\cal S$) and the Dirac operator ${\scr D}$ for our model in the units where $c=e=\hbar=1$ read
\begin{align}
    &\tilde{\cal Z} \propto \int {\cal D}\Psi {\cal D}\overline{\Psi} e^{\tilde{\cal S}}
    , \quad 
    \text{with    }
    \tilde{\cal S}=\tilde{\cal S}_{0}+\tilde{\cal S}_{\rm int}
    \label{eq:action_int}
    ,\\
    & \tilde{\cal S}_{0} = \i \int d^{d} x \overline{\Psi} \gamma^{\mu} \tilde{\scr{ D}}_{\mu} \Psi
    ,\\
      & \tilde{\cal S}_{\rm int} = \int d^{d} x 
      \left(
    -\frac{\lambda_{\mu \nu}^{2}}{2} j^{\mu} j^{\nu}
    -\frac{\lambda_{5,\mu \nu}^{2}}{2} j^{5,\mu} j^{5,\nu}
      \right)
    ,\\
   & \tilde{\slashed{\scr D}} =\gamma^{\mu} \tilde{\scr{ D}}_{\mu} = \gamma^{\mu} \tilde{d}_{\mu} - \i \gamma^{\mu} \tilde{V}_{\mu} -\i \gamma^{\mu}  \gamma^{5} \tilde{W}_{\mu},
\end{align}
 with $\tilde{d}_{\mu}= f_{\mu}^{\nu} \partial_{\nu}$. The mapping between elements of the unified notation and their counterparts in the Hermitionized and non-Hermitian models is introduced in Table~\ref{tab:convertI}.
The gamma matrices $\gamma^{\mu}$ satisfy the Clifford algebra $\{ \gamma^{\mu}, \gamma^{\nu}\}=2 g^{\mu \nu}$ with the Euclidean metric being $g^{\mu \nu} =- \delta^{\mu \nu}$ and Greek indices run from 1 to $d$. The Hermitian fifth gamma matrix reads $\gamma^{5}=-\prod_{\mu} \gamma^{\mu}$ and $\gamma^{0}=\i \gamma^{d}$ used in obtaining the Dirac adjoint $\overline{\Psi}= \Psi^{\dagger} \gamma^{0}$. The Fermi velocities are elements of the rank $d$ diagonal matrix $M={\rm diag}[v_{1}, \ldots, v_{d}]$ with $v_{d}=1$ and $v_{i \neq d}$ are in general complex-valued Fermi velocities.

\begin{table}
\begin{center}
\begin{tabular}{c c  c  c  c}
\hline
\hline
\\[-0.95em]
$\tilde{\mathcal{S}}$ &  $f_{\mu}^{\nu}$ & $\tilde{A}_{\mu}$ &    $\tilde{\cal F}_{2}$ &  $\tilde{\cal F}_{4}$
    \\[-0.05em]
     \hline
    &&\\[-0.05em]
     $\mathcal{S}_{\rm{h}}$  & $\, \Re[M_{\mu}^{\nu}] \, $   & $\, \quad \Re[M_{\mu}^{\nu} A_{\nu}]\, $  &
     $\, 4 \pi |\Re[v_{\rm f} ]| \,$  & $\, 32 \pi^2 |\det[\Re[M]]| $ \,
    \\[-0.05em]
    &&\\[-0.05em]
     $\mathcal{S}_{\rm{nh}}$ & $\, M_{\mu}^{\nu} \,$ &$\, M_{\mu}^{\nu} A_{\nu}\, $ 
     & $\, 4 \pi \sqrt{\det[B]} \,$ & $\, 32 \pi^2 \sqrt{\det[B]}$
      \\[-0.05em]
    \hline
\hline
\end{tabular}
\end{center}
\caption{\label{tab:convertI} Mapping $f$, $\tilde{A}$ and $\tilde{\cal F}$ from the unified notation into Hermitianized and non-Hermitian notation. $A$ stands for gauge fields $V$ and $W$.  $\tilde{\cal F}$ is presented for $2$ and $4$ dimensions. The matrix $B$ for non-Hermitian models is given in Eq.~\eqref{eq:constB}. The elements of the diagonal matrix $M$ are complex-valued Fermi velocities.
}
\end{table}

The interacting action $\tilde{\cal S}_{\rm int}$ consists of short-range four-fermion interactions between currents~($j^{\mu}= \overline{\Psi} \gamma^{\mu} \Psi$) and chiral currents~($j^{5,\mu}= \overline{\Psi}  \gamma^{\mu}  \gamma^{5} \Psi$) with real-valued interaction strengths $\lambda_{\mu \nu}^{2}=\lambda_{\mu \alpha} \lambda_{\nu}^{\alpha}$ and $\lambda_{5,\mu \nu}^{2}=\lambda_{5,\mu \alpha} \lambda_{5,\nu}^{\alpha}$, respectively~\footnote{We emphasize that merely real-valued interaction strengths can be handled in this work as the key Hubbard-Stratonovich transformation breaks down for complex coefficients.}. The current-current interaction with interaction strengths $\lambda_{\mu \nu}^{2}=\lambda^{2} g_{\mu d} g_{\nu d}$ describes a density-density interaction. Similarly, the interaction between chiral currents with $\lambda_{5,\mu \nu}^{2}=\lambda^{2}_{5} g_{\mu d} g_{\nu d}$ in $d=4$ dimensions embed the spin-spin interaction in its spatial part~\cite{Parhizkar2023}.

The action $\tilde{S}$ respects $U_{A}(1) \times U_{V}(1)$ symmetry classically~\cite{Rylands2021, Parhizkar2023}, where $U_{A(V)}(1)$ denotes the chiral~(vector) symmetry. However, this classical symmetry does not hold in the presence of quantum fluctuations resulting in the emergence of the chiral anomaly. In the following, we present the covariant form of this anomaly using Fujikawa's path integral approach~\cite{Fujikawa1979, Fujikawa1980, Fujikawa1981, Bertlmann1996, Fujikawa2004}.

We start with introducing two auxiliary field $a$ and $s$ which brings the interaction action $\tilde{S}$ in Eq.~\eqref{eq:action_int} into a free-fermion action $\tilde{\cal S}_{a,s}$ through a Hubbard-Stratonovich transformation such that
\begin{align}
   & \int {\cal D}\Psi {\cal D}\overline{\Psi} e^{\tilde{\cal S}}
    =
    \int {\cal D}\Psi {\cal D}\overline{\Psi}{\cal D}a {\cal D}s \, e^{\tilde{\cal S}_{a,s}} \equiv \tilde{\cal Z}_{a,s}
    \label{eq:Z_as}
    ,
    \\
   & \tilde{\cal S}_{a,s} =  \int d^{d} x \left[
    \i \overline{\Psi} \gamma^{\mu} \tilde{\scr{ D}}_{a,s,\mu} \Psi
    + \frac{1}{2} a_{\mu}a^{\mu}  + \frac{1}{2} s_{\mu}s^{\mu} 
    \right]
    \label{eq:S_as}
    ,\\
    & \tilde{\scr{ D}}_{a,s,\mu} =  \tilde{d}_{\mu} -\i \tilde{V}_{\mu} - \i \gamma^{5} \tilde{W}_{\mu}  -\i {\lambda}_{\mu \nu} a^{\nu}-\i {\lambda}_{5 \mu \nu} \gamma^{5} s^{\nu} .
\end{align}
Integration over $a$ and $s$ fields in the above equations reproduces $\tilde{\cal S}$ in Eq.~\eqref{eq:action_int}. Performing an infinitesimal chiral transformation wth angle $\beta$ on the spinor $\Psi_{\rm rot}=\exp[-\i \gamma^{5} \beta(x)] \Psi$, keeps the action invariant but gives rise to an anomalous term due to a change in the Jacobian of the path integral measure such that
\begin{align}
    &{\cal D} \Psi_{\rm rot} {\cal D} \overline{\Psi}_{\rm rot}
    =e^{\i \int d^{d} x \beta(x) \tilde{{\cal A}}_{a,s}^{5}}
    {\cal D} \Psi {\cal D} \overline{\Psi}.
    \label{eq:S5A5}
\end{align}

To evaluate this anomalous contribution, we express $\Psi$ and $\overline{\Psi}$ in terms of eigenbasis of the Hermitian Laplacian operators 
$\tilde{\slashed{\scr{D}}}_{a,s} \tilde{\slashed{\scr{D}}}^{\dagger}_{a,s}$ and
$\tilde{\slashed{\scr{D}}}^{\dagger}_{a,s} \tilde{\slashed{\scr{D}}}_{a,s}$~\cite{Kikuchi1992, Rylands2021} 
and employ the Heat-kernel method~\cite{Nakahara1990, Bertlmann1996, Fujikawa2004} to regularize the divergent sum in the exponent of the Jacobian; see Ref.~\cite{Sayyad2022ca} for further details on a similar approach. Introducing $\overline{V}_{\mu} = \tilde{V}_{\mu} + \lambda_{\mu \nu} a^{\nu}$ and $\overline{W}_{\mu}=\tilde{W}_{\mu} + \lambda_{5\mu \nu} s^{\nu}$, $\tilde{\cal A}_{a,s}^{5}$ in $d=2$ dimensions and up to the first order in fields casts
 \begin{equation}
 \tilde{{\cal A}}_{a,s}^{5}  =\frac{ -\varepsilon^{\mu \nu} }{\tilde{\cal F}_{2}} 
       \left[
\i (
  \tilde{F}_{\mu \nu}[ \overline{V}^{\dagger}] 
         -
  \tilde{F}^{\dagger}_{\mu \nu}[ \overline{V}] 
  )
 \right]
.\label{eq:A5_1p1_mt_Vbar} 
 \end{equation}
 In $d=4$ dimensions and up to the second-order in the fields, $\tilde{\cal A}_{a,s}^{5}$ reads
    \begin{align}
\tilde{\cal A}_{a,s}^{5}
 =&
   \frac{   \varepsilon^{\mu \nu \eta \zeta}}{\tilde{\cal F}_{4}} 
     \left[
\tilde{F}_{\mu \nu}[ \overline{V}^{\dagger}] 
\tilde{F}_{\eta \zeta}[ \overline{V}^{\dagger}] 
+
 \tilde{F}^{\dagger}_{\mu \nu}[ \overline{V}] 
\tilde{F}^{\dagger}_{\eta \zeta}[ \overline{V}] 
  \right.
    \nonumber \\
  &
  \left.
+
\tilde{F}^{\dagger}_{\mu \nu} [\overline{W}]
  \tilde{F}^{\dagger}_{\eta\zeta} [\overline{W}]
  +
 \tilde{F}_{\mu \nu} [\overline{W}^{\dagger}]
 \tilde{F}_{\eta \zeta} [\overline{W}^{\dagger}]
 \right]
.\label{eq:A5_3p1_Vbar}
\end{align}
Here, $\tilde{F}_{\mu \nu} [\overline{A}]= \tilde{d}_{\mu} \overline{A}_{\nu}- \tilde{d}_{\nu} \overline{A}_{\mu} $, and $\tilde{F}^{\dagger}_{\mu \nu} [\overline{A}]= \tilde{d}^{\dagger}_{\mu} \overline{A}_{\nu}- \tilde{d}^{\dagger}_{\nu} \overline{A}_{\mu} $ with $\tilde{d}_{\mu}^{\dagger} = - f^{ * \nu }_{\mu} \partial_{\nu}$. $\tilde{{\cal F}}_{d}$ for Hermitionized and non-Hermitian models in $d=2,4$ dimensions are presented in Table.~\ref{tab:convertI}, where it is written in terms of the determinant of a matrix $B$ given in Euclidean space, with matrix elements
  \begin{equation}
 B^{\alpha \beta}= \delta^{\mu \nu}
 f_{\mu}^{*\alpha} f_{\nu}^{\beta}
 -
  \frac{1}{2} [ \gamma^{ \mu},  \gamma^{ \nu}]
\frac{f_{\mu}^{*\alpha} f_{\nu}^{\beta} 
- f_{\mu}^{\alpha} f_{\nu}^{*\beta}   
}{2}. \label{eq:constB}
 \end{equation}

 Carrying out the same steps for an infinitesimal vector transformation $\Psi_{\rm rot} = \exp[-\i \kappa(x)] \Psi$ with $  {\cal D} \Psi_{\rm rot} {\cal D}\bar{\Psi}_{\rm rot}
 =\exp \left[ {\i \int {\rm d}^{d} x \kappa(x) \tilde{\cal A}_{a,s}} \right] {\cal D} \Psi {\cal D}\bar{\Psi}$ in $d=2$ dimensions and up to the first order in the fields, results in
   \begin{align}
 \tilde{\cal A}_{a,s}= & \frac{-\varepsilon^{\mu \nu} }{\tilde{\cal F}_{2}} 
     \left[ 
 \i \big( 
 \tilde{F}^{\dagger}_{\mu \nu} [\overline{W}]
  -  \tilde{F}_{\mu \nu} [\overline{W}^{\dagger}]
      \big)
 \right]
    ,\label{eq:A_1p1_mt_Vbar}
 \end{align}
 and in $d=4$ dimensions and up to the second order in fields, $ \tilde{\cal A}_{a,s}$ reads
 \begin{align}
\tilde{\cal A}_{a,s} 
 &=
  \frac{- \varepsilon^{\mu \nu \eta \zeta} }{\tilde{\cal F}_{4}} 
   \left[ 
 \tilde{F}_{\mu \nu}[ \overline{V}^{\dagger}] 
 \tilde{F}_{\eta \zeta} [\overline{W}^{\dagger}]
+
\tilde{F}^{\dagger}_{\mu \nu}[ \overline{V}]  
 \tilde{F}^{\dagger}_{\eta \zeta} [\overline{W}] 
 \right].\label{eq:A_3p1}
 \end{align}
When $\lambda_{\mu \nu} $ and $ \lambda_{5 \mu \nu}$ are zero, $\tilde{\cal A}_{a,s}$ and $\tilde{\cal A}_{a,s}^{5}$ reproduce the results of Ref.~\cite{Sayyad2022ca}.

 Combining all results, the rotated generalized action in Eq.~\eqref{eq:S_as} under the vector and chiral transformation casts
 \begin{align}
\label{eq:currentrot}
    \tilde{\cal S}_{a,s}^{{\rm rot }} -\tilde{\cal S}_{a,s} =& -  \int {\rm d}^{2} x   \Big[-\beta(x)  \tilde{d}_{\mu}  j_{a,s}^{5,\mu} 
    - \kappa(x) 
      \tilde{d}_{\mu}  j_{a,s}^{\mu} \Big].
\end{align}
Enforcing the invariance of the partition function $\tilde{\cal Z}_{a,s}$ in Eq.~\eqref{eq:Z_as} under the vector and chiral transformations, results in satisfying $\tilde{\cal A}_{a,s}^{5}= -\i \tilde{d}_{\mu}  j_{a,s}^{5,\mu} $ and $\tilde{\cal A}_{a,s}= -\i \tilde{d}_{\mu}  j_{a,s}^{\mu}$. To obtain the anomalous relations for the interacting model, we should shift the auxiliary fields by their on-shell values as $a_{\mu}\rightarrow a_{\mu} - \lambda_{\mu \alpha} j^{\alpha}$ and $s_{\mu}\rightarrow s_{\mu} - \lambda_{5\mu \alpha} j^{5,\alpha}$ and integrate over Hubbard-Stratonovich fields $a$ and $s$~\cite{Parhizkar2023}. The subsequent anomalous equation in the Euclidean space in $d=2$ dimensions is
 \begin{equation}
 \tilde{d}_{\mu}  j^{5,\mu} =\frac{ \varepsilon^{\mu \nu} }{\tilde{\cal F}_{2}} 
       \left(
  \tilde{F}_{\mu \nu}[ \tilde{V}^{\dagger}] 
         -
  \tilde{F}^{\dagger}_{\mu \nu}[ \tilde{V}] 
  -
   4 \Re[f_{\mu}^{\eta}] \partial_{\eta}[ \lambda_{\nu \alpha}^{2} j^{\alpha}]
 \right)
,\label{eq:dj5_2} 
 \end{equation}
 and in $d=4$ dimensions reads
    \begin{align}
 \tilde{d}_{\mu}  j^{5,\mu}
 &=
   \frac{   \varepsilon^{\mu \nu \eta \zeta}}{\tilde{\cal F}_{4}} 
     \left(
\tilde{F}_{\mu \nu}[ \tilde{V}^{\dagger}] 
\tilde{F}_{\eta \zeta}[ \tilde{V}^{\dagger}] 
+
 \tilde{F}^{\dagger}_{\mu \nu}[ \tilde{V}] 
\tilde{F}^{\dagger}_{\eta \zeta}[ \tilde{V}] 
  \right.
    \nonumber \\
  &
+
\tilde{F}^{\dagger}_{\mu \nu} [\tilde{W}]
  \tilde{F}^{\dagger}_{\eta\zeta} [\tilde{W}]
  +
 \tilde{F}_{\mu \nu} [\tilde{W}^{\dagger}]
 \tilde{F}_{\eta \zeta} [\tilde{W}^{\dagger}]
 \nonumber \\
 &
- 4 
\tilde{d}_{\eta} [ \lambda^{2}_{\zeta \xi} j^{\xi}] 
\,
\tilde{F}_{\mu \nu}[ \tilde{V}^{\dagger}] 
-4
\tilde{d}_{\eta }^{\dagger} [ \lambda^{2}_{\zeta \xi} j^{\xi}] 
\,
 \tilde{F}^{\dagger}_{\mu \nu}[ \tilde{V}] 
 \nonumber \\
 &
- 4
\tilde{d}_{\eta} [ \lambda^{2}_{5,\zeta \xi} j^{5,\xi}] 
\,
\tilde{F}_{\mu \nu}[ \tilde{W}^{\dagger}] 
-4
\tilde{d}_{\eta }^{\dagger} [ \lambda^{2}_{5,\zeta \xi} j^{5,\xi}] 
\,
 \tilde{F}^{\dagger}_{\mu \nu}[ \tilde{W}] 
  \nonumber \\
&
   +
   8 \Re[f_{\mu }^{ \kappa} f_{\eta }^{ \iota}]
   \partial_{\kappa} [\lambda^{2}_{\nu \alpha} j^{\alpha}] 
    \partial_{\iota} [\lambda^{2}_{\zeta \alpha} j^{\alpha}] 
   \nonumber \\
 &
   \left.
   +
   8 \Re[f_{\mu }^{ \kappa} f_{\eta }^{ \iota}]
   \partial_{\kappa} [\lambda^{2}_{5,\nu \alpha} j^{5,\alpha}] 
    \partial_{\iota} [\lambda^{2}_{5,\zeta \alpha} j^{5,\alpha}] 
 \right)
.\label{eq:dj5_4}
\end{align} 
When the gauge field $V$ is real, $W$ is absent and $M= \id_{d \times d}$, the above $\tilde{d}_{\mu}  j^{5,\mu}$ are in agreement with the Hermitian results~\cite{Rylands2021, Parhizkar2023}. We note that in the absence of interactions, the above relations reproduce results of Ref.~\cite{Sayyad2022ca}.

In the Minkowski space, the divergence of chiral currents in Eqs.~\eqref{eq:dj5_2} and \eqref{eq:dj5_4} read
\begin{align}
\tilde{d}_{\mu} j^{5,\mu}  =
& \frac{  2}{\tilde{\cal F}_{2}} 
     \Big(   \tilde{E}_{1}^{\dagger} + \tilde{E}_{1}
\nonumber \\
&
-4 \Re[v_{\mu}] \delta^{\mu \eta} \partial_{\eta}[ \lambda_{\nu \alpha}^{2} j^{\alpha}]
\Big)
\, \,
\text{
 in $d=1+1$} 
,\label{eq:dJ5_1p1}
\\
\tilde{d}_{\mu} j^{5,\mu} = & 
 \frac{   8 }{\tilde{\cal F}_{4}} 
     \left(   \overline{\bolds{E}}^{\dagger}\cdot \overline{\bolds{B}}^{\dagger} + \overline{\bolds{E}}^{5\dagger}\cdot \overline{\bolds{B}}^{5\dagger} 
     \right.
\nonumber \\
& \qquad 
\left.
+\overline{\bolds{E}}\cdot \overline{\bolds{B}}+ \overline{\bolds{E}}^{5}\cdot \overline{\bolds{B}}^{5} 
\right) 
\,
\, \, 
\text{ in $d=3+1$}.
\label{eq:dJ5_3p1}
\end{align}
Here, the generalized electric and magnetic fields cast
\begin{align}
    \overline{E}_{\mu}
    &=v_{\mu}^{ *}  \delta^{\mu \iota} \tilde{E}_{\iota}
    -
    \big(
\partial_{0} {\lambda}^{2}_{\mu \alpha} 
- v_{\mu}^{*} \delta^{\mu \iota} \partial_{\iota} {\lambda}^{2}_{0 \alpha} 
    \big) j^{\alpha}
    \label{eq:Ebar}
    ,\\
        \overline{E}^{5}_{\mu}
    &=v_{\mu}^{ *} \delta^{\mu \iota} \tilde{E}^{5}_{\iota}
    -
    \big(
\partial_{0} {\lambda}^{2}_{5,\mu \alpha} 
- v_{\mu}^{*} \delta^{\mu \iota} \partial_{\iota} {\lambda}^{2}_{5,0 \alpha} 
    \big) j^{5,\alpha}
     \label{eq:E5bar}
       ,\\
        \overline{B}_{\xi \zeta}
    =&v_{\xi}^{ *} v_{\zeta}^{ *} \delta^{\xi \iota} \delta^{\zeta \kappa} \tilde{B}_{\iota \kappa}
    \nonumber \\
    &
    -
     \frac{1}{2} 
    \big(
v_{\xi}^{ *} \delta^{\xi \iota} \partial_{\iota}  {\lambda}^{2}_{\zeta \alpha} 
- v_{\zeta}^{*} \delta^{\zeta \kappa} \partial_{\kappa} {\lambda}^{2}_{\xi \alpha} 
    \big) j^{\alpha}
     \label{eq:Bbar}
    ,\\
        \overline{B}^{5}_{\xi \zeta}
    =&v_{\xi}^{ *} v_{\zeta}^{ *} \delta^{\xi \iota} \delta^{\zeta \kappa} \tilde{B}^{5}_{\iota \kappa}
    \nonumber \\
    &
    -
  \frac{1}{2}  \big(
v_{\xi}^{ *} \delta^{\xi \iota} \partial_{\iota}  {\lambda}^{2}_{5,\zeta \alpha} 
- v_{\zeta}^{*} \delta^{\zeta \kappa} \partial_{\kappa} {\lambda}^{2}_{5,\xi \alpha} 
    \big) j^{5,\alpha}
     \label{eq:B5bar}
    .
\end{align}
The complex electric fields are also given by $\tilde{E}_{j}= (\exp[2 \i \phi_{j} ] \partial_{t} V_{j} -  \partial_{j} V_{0})$, $\tilde{E}^{5}_{j}= (\exp[2 \i \phi_{j} ] \partial_{t} W_{j} -  \partial_{j} W_{0})$ with $i,j,k \neq t$. Similarly, the complex magnetic fields cast $\tilde{B}^{i} =\varepsilon^{ijk}\tilde{B}_{jk}$ and  $\tilde{B}^{5,i} =\varepsilon^{ijk} \tilde{B}_{jk}^{5}$ with $\tilde{B}_{jk}= \exp[2 \i \phi_{k}] \partial_{j} V_{k} -\exp[2 \i \phi_{j}]  \partial_{k} V_{j} $ and $\tilde{B}_{jk}^{5}= \exp[2 \i \phi_{k}] \partial_{j} W_{k} -\exp[2 \i \phi_{j}]  \partial_{k} W_{j}$. The phase $\phi_{j}$ with $j$ being a spatial index satisfies $\exp[\i \phi_{j}]=v_{j}/ |v_{j}|$. 

Our main results in Eqs. \eqref{eq:dJ5_1p1} and \eqref{eq:dJ5_3p1} retain the general forms of chiral anomaly, namely $\tilde{d}_{\mu}j^{5,\mu} \propto E$ in $(1+1)$ dimensions and $\tilde{d}_{\mu}j^{5,\mu} \propto \bolds{E}. \bolds{B}$ in $(3+1)$ dimensions, in Hermitian systems~\cite{Fujikawa1979, Bertlmann1996, Rylands2021, Parhizkar2023}. We emphasize that anomalous relations carry additional terms originating from complex Fermi velocities and complex gauge fields not present in Hermitionized anomalous equations. Note also that the terms proportional to interaction strengths can be written as total derivatives of physical quantities. The prefactors of $\tilde{\cal F}_{d}$ in non-Hermitian systems with complex Fermi velocities also differ from their counterparts in Hermitionized models.

The complex generalized fields in Eqs.~(\ref{eq:Ebar}, \ref{eq:E5bar}, \ref{eq:Bbar}, \ref{eq:B5bar}) can be viewed as complex background fields screened by interactions between currents and densities of spinors. These screened fields are then responsible for breaking the chiral symmetry and giving rise to the chiral anomaly. In other words, anomalous relations in Eqs. \eqref{eq:dJ5_1p1} and \eqref{eq:dJ5_3p1} account for violating the chiral symmetry by the gauge fields $(V, W)$ and also by the induced contributions from interactions between different constituent of our systems.

Considering $(1+1)$-dimensional systems with real Fermi velocities and $\lambda^{\mu \nu}=\lambda^{2} \delta^{\mu \nu}$ with $\lambda$ being a constant, we rewrite Eq.~\eqref{eq:dJ5_1p1} as
\begin{align}
\tilde{d}_{\mu} j^{5,\mu}  =
& \frac{   1}{1  + 4 \lambda^2/\tilde{\cal F}_{2} } \frac{   2}{\tilde{\cal F}_{2}}
     \left(   \tilde{E}_{1}^{\dagger} + \tilde{E}_{1}
\right)
\label{eq:dj5_Bz_1p1}
,
\end{align}
where we use the relation between the chiral and vector currents $j^{5,\mu} = \epsilon^{\mu \nu} j_{\nu}$ to obtain the above relation. In the absence of the axial $W$ field in $(3+1)$ dimensions, keeping $M \in \mathbb{R}$, $\lambda^{\mu \nu}=\lambda^{2} \delta^{\mu \nu}$ and setting $\bolds{E} = E_{z} \hat{z} $ and $\bolds{B} = B_{z} \hat{z} $ results in
\begin{align}
   \tilde{d}_{\mu} j^{5,\mu}  =
& \frac{   1}{1  + \frac{8 \lambda^2 (\tilde{B}_{z} + \tilde{B}^{\dagger}_{z})}{\tilde{\cal F}_{4} } } \frac{   8}{\tilde{\cal F}_{4} }
     \left(   \tilde{E}_{z}^{\dagger} \tilde{B}_{z}^{\dagger} 
     +
     \tilde{E}_{z} \tilde{B}_{z}
\right)
\label{eq:dj5_Bz_3p1}
,
\end{align}
with the relation $\varepsilon^{12 \mu \nu } j_{\nu}=j^{5,\mu}$. We note that the current-current contributions, e.g., the last two terms in Eq.~\eqref{eq:dj5_4}, do not appear in the above relation. This is due to the translational and rotational symmetry on the $x-y$ plane, with fields along the $z$ directions. 
We can interpret our results in Eqs.~\eqref{eq:dj5_Bz_1p1} and \eqref{eq:dj5_Bz_3p1} as renormalization of $(\bolds{E},\bolds{B})$ fields by interactions. These equations coincide with the Hermitian results in $(1+1)$~\cite{Shei1972, Parhizkar2023} and $(3+1)$~\cite{Rylands2021, Parhizkar2023} dimensions upon imposing $(E, B)$ to be real fields.

\paragraph*{Physical consequences in non-Hermitian Weyl semimetals.}
To obtain physical phenomena stemming from the non-Hermitian chiral anomaly in interacting systems, we proceed with presenting the Chern-Simons description of our $(3+1)$ dimensional model in the absence of the axial field $W$ and without interactions between chiral currents. In this case, the change of action $\tilde{S}$ under an infinitesimal rotation $\beta$ reads
\begin{align}
    \tilde{\cal S}_{\rm rot} - \tilde{\cal S} = \int d t d^{3} x \beta(t,x) [\tilde{d}_{\mu} j^{5,\mu} - \tilde{A}_{5}],
\end{align}
where the first term in the r.h.s. of the above relation is due to the classical shift of the action and the anomalous second term~($\tilde{S}^{5}$) accounts for the change of measure and should satisfy $\tilde{A}_{5}=\tilde{d}_{\mu} j^{5,\mu} $ as we discussed in the previous section. The anomaly-induced action after performing integration by parts and neglecting a total derivative term can be rewritten as the Chern-Simons action in the Minkowski spacetime as
\begin{align}
    \tilde{S}^{5}[\beta] = & - \int d t d^{3} x
     \frac{8\varepsilon^{\mu \nu \eta \zeta}}{\tilde{F}_{4}} 
     \left[
     \tilde{d}_{\eta} \tilde{j}_{\zeta} \tilde{d}_{\mu} \beta \tilde{V}^{\dagger}_{\nu}
+
     \tilde{d}^{\dagger}_{\eta} \tilde{j}_{\zeta} \tilde{d}^{\dagger}_{\mu} \beta \tilde{V}_{\nu}
     \right]
           \nonumber \\
     &
     + \int d t d^{3} x
     \frac{8\varepsilon^{\mu \nu \eta \zeta}}{\tilde{F}_{4}} 
     \Re[f_{\mu }^{ \kappa} f_{\eta }^{ \iota}]
   \partial_{\kappa} \beta \tilde{j}_{\nu} 
    \partial_{\iota} \tilde{j}_{\zeta} 
         \nonumber \\
     &
     +
    \int d t d^{3} x
     \frac{4\varepsilon^{\mu \nu \eta \zeta}}{\tilde{F}_{4}} 
     \tilde{d}_{\mu} \beta \tilde{V}^{\dagger}_{\nu} \tilde{d}_{\eta} \tilde{V}^{\dagger}_{\zeta}
     \nonumber \\
     &+ \int d t d^{3} x
     \frac{4\varepsilon^{\mu \nu \eta \zeta}}{\tilde{F}_{4}} 
     \tilde{d}^{\dagger}_{\mu} \beta \tilde{V}_{\nu} \tilde{d}^{\dagger}_{\eta} \tilde{V}_{\zeta},
\end{align}
where $\tilde{j}_{\mu} =\lambda^{2}_{\mu \alpha} j^{\alpha}$. The associated currents for the above action are then evaluated by summing the functional derivatives of $\tilde{S}^{5}$ with respect to $V$ and $V^{\dagger}$. These currents are given by
\begin{align}
    M_{\nu}^{\alpha} j^{\nu}
    =&
     \frac{8\varepsilon^{\mu \nu \eta \zeta} \partial_{\delta} \beta}{\tilde{F}_{4}} 
     \Re \left[ M_{\mu}^{ \delta} M_{\eta}^{\iota} M_{\zeta}^{ \rho} M_{\nu}^{*\alpha} \partial_{\iota} V_{\rho} \right]
     \nonumber \\
     &
     -\frac{16\varepsilon^{\mu \nu \eta \zeta} \partial_{\delta} \beta}{\tilde{F}_{4}} 
     \Re \left[ M_{\eta}^{\iota} M_{\mu}^{\delta} M_{\nu}^{*\alpha} \right]
     \partial_{\iota} \tilde{j}_{\zeta} .
     \label{eq:jCS}
\end{align}
Imposing $V$ field to be real and $M=\id_{4\times4}$ in Eq.~\eqref{eq:jCS} recovers the Hermitian chiral magnetic effect when $\delta=0$ and the Hermitian anomalous Hall effect with $\delta$ being a spatial index~\cite{Zyuzin2012}. 
We note that the interaction-induced terms, second line in Eq.~\eqref{eq:jCS}, are merely present in nonequilibrium systems~($\partial_{\iota} j_{\nu} \neq 0$ with $\nu \neq \iota =0$) or in inhomogeneous systems with $\partial_{\iota} j_{0} \neq 0$ with $\iota \neq 0$.
Hence, in nonequilibrium or inhomogeneous systems, Eq.~\eqref{eq:jCS} describes the interacting non-Hermitian chiral magnetic effect when the temporal component of $\beta$~($\delta=0$) is nonzero and when a spatial component of $\beta$ with $\delta \in \{1,2,3\}$ is nonvanishing, Eq.~\eqref{eq:jCS} results in the interacting non-Hermitian anomalous Hall effect.

 $\partial_{0}\beta$ and $\partial_{\delta}\beta$ are related to the complex-valued energy and spatial separation of the Weyl nodes in non-Hermitian Weyl semimetals~\cite{Sayyad2022ca}. We emphasize that these Weyl points should be non-defective degeneracies. This is because the coalescence of eigenvectors in defective degeneracies prevents introducing a well-defined basis to express the Laplacian operators for Fujikawa's path integral method. As all non-defective degeneracies are symmetry-protected~\cite{Sayyad2022pro}, respecting their underlying symmetry maintains nodal intersections and offers platforms to realize non-Hermitian chiral anomalies. One approach for constructing non-Hermitian Weyl semimetals is the effective non-Hermitian description of open quantum systems. Starting from a Hermitian model for Weyl semimetals and allowing the coupling of this system with external environments, e.g., by a Boltzmann factor~(see Ref.~\cite{Nuske2020} for an example), results in an effective Hamiltonian whose imaginary part of the spectrum is nonpositive.

\paragraph*{Conclusion and outlook.}
In conclusion, we have presented how non-Hermitian chiral anomalies for massless fermions with complex Fermi velocities coupled to complex background fields are modified in the presence of four-particle interactions. We have shown that anomalous relations cast the same form as the chiral anomaly in Hermitian (non)interacting systems. Despite the similar structure, the embedded terms in the presented non-Hermitian chiral anomaly exceed those in the Hermitionized model. Our results show that nonperturbative interaction corrections to chiral anomalies are nonvanishing in nonequilibrium or inhomogeneous systems. This can be seen in the presented non-Hermitian chiral magnetic effect and non-Hermitian anomalous Hall effect. 

An experimental study on light-driven anomalous Hall effect in graphene reported that the Hall conductance is unquantized, despite the theoretical expectation of quantized conductivity in Hermitian systems~\cite{Mciver2020}. Theoretical efforts to explain this observation revealed the essential roles played by out-of-equilibrium and dissipative properties of the experimental system~\cite{Sato2019, Nuske2020}. As these two factors are well incorporated within our theory and measuring additional, terms proportional to currents, in Eqs.~(\ref{eq:dj5_2}, \ref{eq:dj5_4}) are experimentally feasible, we expect to find signatures of our findings in light-driven Weyl semimetals with a similar experimental setup as in Ref.~\cite{Mciver2020}. In addition, combining circuits to explore real-time chiral dynamics~\cite{Kharzeev2020} with algorithms to simulate open quantum systems~\cite{Kamakari2022} may also pave the way to realize our findings digitally.

Finally, extending these results to understand the parity-like anomaly~\cite{Burkov2018, Matsushita2020}, the axial-torsional anomaly~\cite{Ferreiros2019} and axial–gravitational anomaly~\cite{Gooth2017, Chernodub2021} in non-Hermitian interacting systems is also of interest which we leave for future studies.

\paragraph*{Acknowledgement.} I thank A. Parhizkar for the clarifications on reproducing some of the results in Ref.~\cite{Parhizkar2023}. 
Fruitful discussions with K. Landsteiner and helpful comments on the manuscript from A.~G. Grushin and P. Surwoka are gratefully acknowledged.

\bibliography{NHAnomaly.bib}
\end{document}